\documentclass{PoS}

\title{Relativistic point-form approach to \\
hadron properties}

\ShortTitle{Point-form approach}

\author{\speaker{Willibald Plessas}
        \\
        Institute of Physics, University of Graz,
        Universit\"atsplatz 5, A-8010 Graz, Austria\\
        E-mail: \email{plessas@uni-graz.at}}

\abstract{
We present a review of the description of hadron properties along an invariant mass operator in the point form of Poincar\'e-invariant relativistic dynamics. The quark-quark interaction is furnished by a linear confinement, consistent with the QCD string tension, and a hyperfine interaction derived from Goldstone-boson exchange. The main advantage of the point-form approach is the possibility of calculating manifestly covariant observables, since the generators of Lorentz transformations remain interaction-free. We discuss the static properties of the mass-operator eigenstates, such as the invariant mass spectra of light- and heavy-flavor baryons, the characteristics of the eigenstates in terms of their spin, flavor, and spatial dependences as well as their classification into spin-flavor multiplets. Regarding dynamical observables we address the electroweak structures of the nucleon and hyperon ground states, including their electric radii, magnetic moments as well as axial charges, and in addition 
a recently derived microscopic description of the $\pi NN$ as well as $\pi N\Delta$ interaction vertices. Except for hadronic resonance decays, which are not addressed here due to space
limitations, all of these observables are obtained in good agreement with existing phenomenology. Relativistic (boost) effects are generally sizable. We conclude that low-energy hadrons can be well described by an effective theory with a
finite number of degrees of freedom, as long as the symmetries of low-energy quantum chromodynamics (spontaneously broken chiral symmetry) as well as special relativity (Poincar\'e invariance) are properly taken into account. The latter requirement is particularly well
and efficiently met in the point-form approach.}

\FullConference{Light Cone 2010 - LC2010\\
		June 14-18, 2010\\
		Valencia, Spain}

\begin{document}

\section{Motivation for the relativistic constituent-quark model}

Up till now all attempts of solving quantum chromodynamics (QCD) are still limited.
They are either constrained to certain energy domains (such as perturbative QCD), have
to resort to truncations (such as effective field theories or functional approaches),
or face severe numerical limitations (such as lattice QCD). Over the past decades we
have also learnt, however, that it is not absolutely necessary to take into account
all aspects of QCD in order to achieve a consistent quantitative
description of hadron properties,
especially in the domain of low and intermediate energies. Rather it has turned out
sufficient to include into a theory the relevant/active degrees of freedom. Most
reliably these degrees of freedom can be identified by considering the symmetries
governing a particular system. With regard to hadrons (at low energies) such symmetries
are certainly suggested by the spontaneous breaking of chiral symmetry (SB$\chi$S) of
QCD in the low-energy domain and by special relativity, i.e. invariance under
Poincar\'e transformations. In other words, the dynamics we employ for the description
of hadrons has to fulfill (at least) the constraints following from these
invariance conditions.

From this point of view, we may formulate the relativistic constituent-quark model
(RCQM). It describes hadrons as few-quark systems along a relativistically invariant,
interacting mass operator $\hat M=\hat M_{\rm free}+\hat M_{\rm int}$.
In this way, we adhere to a Hamiltonian theory restricted to a finite
number of degrees of freedom and not a field theory. As a consequence it is certainly
limited to a restricted (energy) domain. As soon as further degrees of freedom appear
beyond the ones initially assumed, the theory is no longer adequate and has to be
extended to include them too; this can be done in principle but in the end requires
considerably more computational efforts. The principal advantages of such
a Hamiltonian approach are that
\begin{itemize}
\vspace{-2mm}
\item the theoretical framework is rigorously established, as a Poincar\'e-invariant
      quantum theory formulated on a given Hilbert space, and
\vspace{-2mm}
\item the corresponding dynamical equations (specifically the eigenvalue problem of
      the mass operator) can be solved to any desired accuracy, as long as a
      reasonable finite number of degrees of freedom are involved.
\end{itemize}

In this spirit, we view hadrons -- and possibly other objects allowed by QCD -- as
consisting of a finite number of constituents interacting by dynamics that strictly
conform to the symmetry constraints from the observed/given invariances.
       
\section{The Goldstone-boson exchange RCQM}

In this contribution we are essentially concerned with baryon properties at low and
intermediate energies. In this context we use primarily the RCQM that assumes baryons
to consist of three confined constituent quarks whose dynamics is mediated by
Goldstone-boson exchange (GBE); the corresponding predictions will be compared also with
other modern RCQMs. The main idea of the GBE RCQM is that the constituents as well
as the exchange bosons are generated by the SB$\chi$S of QCD, which means,
specifically in the three-flavor case, the breaking of
$SU(3)_L\times SU(3)_R\rightarrow SU(3)_V$. Consequently, we have constituent quarks
as quasiparticles with a dynamical mass and Goldstone bosons as the generators of the
$SU(3)_V$ transformations. The latter give rise to the forces between the confined
constituent quarks and materialize themselves through pseudoscalar mesons. The effective
interaction Lagrangian is thus given by
\begin{equation}
{\cal L_{\rm int}}\sim ig\bar\psi\gamma_5\vec\lambda^F\cdot \vec\phi \psi \,,
\label{Lint}
\end{equation}    
where $\psi$ are the fermion (constituent quark) fields coupled by the (Goldstone)
boson fields $\vec\phi$ through a pseudoscalar-type coupling involving the $SU(3)$
flavor matrices $\vec\lambda^F$ with a strength $g$.

After such type of dynamics had been suggested by Glozman and Riska for baryons at low
energies~\cite{gloris}, the Graz group constructed a first nonrelativistic constituent
quark model whose hyperfine interaction was deduced from the interaction in eq.~(\ref{Lint})
and led to a quark-quark potential with explicit flavor dependence~\cite{gpp}. Soon it was realized that a nonrelativistic quark model is by no means adequate and one has to work
in a relativistic framework. This resulted in the so-called pseudoscalar
Goldstone-boson exchange relativistic constituent quark model
(psGBE RCQM)~\cite{Glozman:1997ag}. It initially relied only on the spin-spin part of
the GBE interaction but was later on extended to include in addition all central, tensor,
and spin-orbit interactions (EGBE RCQM)~\cite{Glantschnig:2004mu}. For the details of
the GBE RCQMs we refer to the original papers.

\section{Baryon spectroscopy}

\begin{figure}[b]
\begin{center}
\includegraphics[clip=,width=10cm]{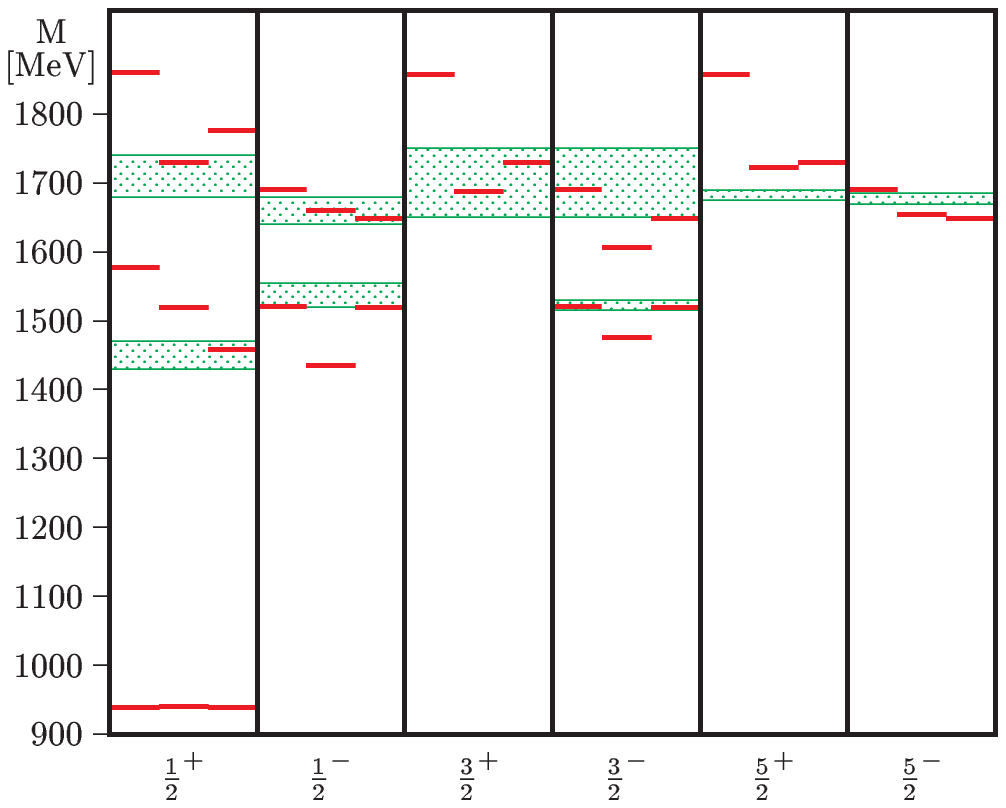}
\end{center}
\vspace{-4mm}
\caption{Nucleon excitation spectra of three different types of RCQMs. In each column
of definite $J^P$ the left horizontal lines represent the results of the relativistic
variant of the Bhaduri-Cohler-Nogami OGE RCQM~\protect\cite{Theussl:2000sj}, the
middle ones of the II RCQM (Version A)~\protect\cite{Loring:2001kx}, and the right ones
of the psGBE RCQM~\protect\cite{Glozman:1997ag}. The shadowed boxes give the experimental
data with their uncertainties after the PDG~\protect\cite{Nakamura:2010zzi}.}
\label{Nspectr}
\vspace{-3mm}
\end{figure}

\begin{figure}[h]
\begin{center}
\includegraphics[width=10cm]{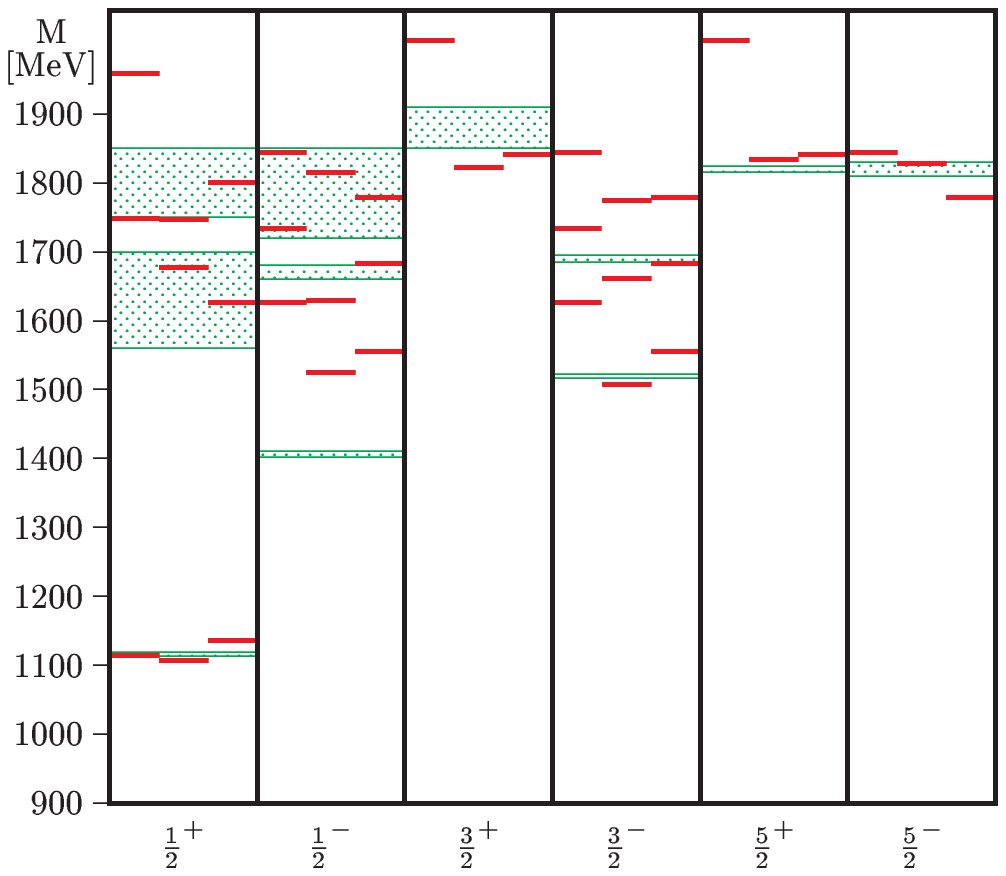}
\end{center}
\vspace{-4mm}
\caption{$\Lambda$ excitation spectra. Same caption as in fig.~\protect\ref{Nspectr}.}
\label{Lspectr}
\vspace{-2mm}
\end{figure}

The remarkable features of the GBE dynamics for constituent quarks are already
seen in the baryon excitation spectra, where for the first time the right
level orderings of positive- and negative-parity resonances could be
reproduced simultaneously in the $N$ and $\Lambda$ spectra~\cite{Glozman:1997fs}.
Obviously this achievement is due to the specific flavor dependence in the pertinent hyperfine potential. Such a behaviour is much in contrast to other recent RCQMs,
like the one-gluon-echange (OGE) RCQM~\cite{Theussl:2000sj} or the one whose hyperfine
potential is deduced from instanton-induced (II) interactions~\cite{Loring:2001kx}, cf.
the comparisons in figs.~\ref{Nspectr} and~\ref{Lspectr}; for the complete baryon
excitation spectra see the original references given above.

The action of the hyperfine forces in the GBE RCQMs is explained in detail in
ref.~\cite{Glozman:1997fs}. There it is also shown, how the level orderings come about,
as one steps out from the case with the confinement interaction only and gradually
increases the size of the meson-quark coupling until its final strength is reached,
which is the value deduced from the pion-nucleon coupling constant via the
Goldberger-Treiman relation. Later on such a behaviour of the level shifts of
positive- and negative-parity resonances has also been found in lattice-QCD calculations,
when one approaches the chiral limit (see, e.g., ref.~\cite{Mathur2005}).

From fig.~\ref{Lspectr} also a severe shortcoming of all the contemporary RCQMs that rely
on three-quark configurations is evident, namely, the failure of reproducing the
$\Lambda$(1405) resonance. Obviously, for this particular case additional degrees of
freedom, specifically the coupling to the $K-N$ channel, are needed. To our knowledge
this has not yet been satisfactorily achieved with a relativistically invariant mass
operator or in any other relativistic framework.

\section{Electroweak structures of baryon ground states}

Once we have solved the eigenvalue problem of the invariant mass operator, we are also
equipped with the eigenstates of the baryon ground states and resonances. They can be
further employed for the calculation of baryon reactions. In this section we consider
the electromagnetic and weak form factors of the baryon ground states.

First of all it is interesting to look at the spatial probability distributions of
these mass-operator eigenstates in the rest frame. Corresponding pictures can be
found in ref.~\cite{Melde:2008yr}. They exhibit particular structures typical for the
classification of ground states and resonances into spin-flavor multiplets. On the
basis of these spatial structures and the strong-decay patterns of the pertinent
resonances a modified and extended classification scheme of baryons into $SU(3)_F$
flavor multiplets has been developed~\cite{Nakamura:2010zzi}. An aspect worth to
be emphasized is the relatively dense probability distribution at short distances
in configuration space. For instance, in case of the $N$ ground state it is maximal
at magnitudes of the Jacobi coordinates of $\xi\approx\eta\approx 0.3$ fm. The
probability distributions for all ground states furthermore exhibit essentially
spherical symmetry and do not show any pronounced quark-diquark
structures~\cite{Melde:2008yr}.
Of course, with regard to certain subtle observables, e.g.,
the electric neutron form factor (see below), small but important mixed-symmetric
spatial configurations in the wave functions are crucially important. 
 
For the calculation of baryon reactions, here specifically the electroweak form
factors as well as electric radii and magnetic moments, one has to know how to
boost the eigenstates. This task can most efficiently be carried out for general
Lorentz transformations in the point form of Poincar\'e-invariant quantum mechanics,
since in this case the generators of rotations and boosts are interaction-free.
Thereby one can calculate manifestly covariant observables.

The results obtained so far for electroweak observables of baryons in the point-form
approach have all been calculated with electromagnetic and axial currents according
to the so-called point-form spectator-model
(PFSM)~\cite{Melde:2004qu}. These approximative currents are defined
by the exchange boson coupling to only one of the constituent quarks, while the
momentum is transferred to the baryon as a whole. Consequently, they amount to effective
many-body currents~\cite{Melde:2007zz}.

\begin{figure}[h]
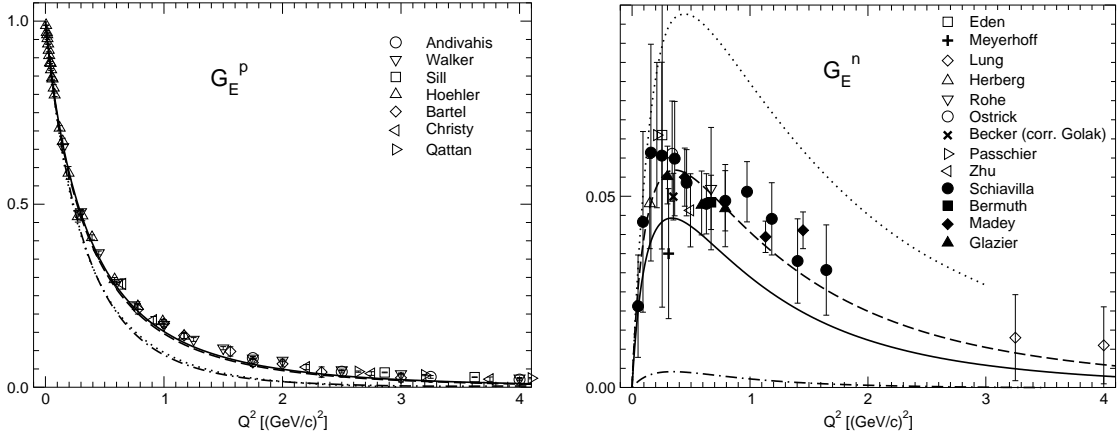

\begin{center}
\includegraphics[clip=,width=7.1cm]{gep_RCQMs.eps} \hspace{4mm}
\includegraphics[clip=,width=7.1cm]{gen_RCQMs.eps}
\end{center}
\vspace{-4mm}
\caption{Predictions of the psGBE RCQM for the electric proton and neutron form factors
as calculated along the PFSM (solid lines) in comparison to the results from the
Bhaduri-Cohler-Nogami OGE RCQM calculated also with the PFSM (dashed lines) and the
results from the II RCQM calculated in a Bethe-Salpeter approach (dotted lines).
A comparison is also given to the case with confinement only, i.e. with a spatially
symmetric wave function (dashed-dotted line). Experimental data as specified in the
inserts, see also ref.~\cite{Melde:2007zz}.}
\label{eFFs}
\end{figure}

\begin{figure}[h]
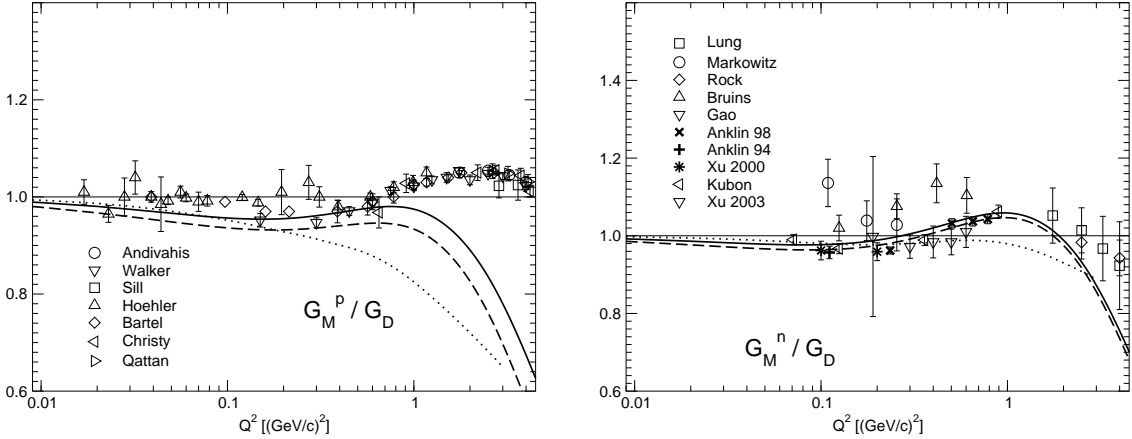

\begin{center}
\includegraphics[clip=,width=7.1cm]{gmpbygd_RCQMs.eps} \hspace{6mm}
\includegraphics[clip=,width=7.1cm]{gmnbygd_RCQMs.eps}
\end{center}
\vspace{-4mm}
\caption{Same comparison as in fig.~\protect\ref{eFFs} but for the magnetic proton
and neutron form factors as represented by ratios to the dipole form.}
\label{mFFs}
\end{figure}

In figs.~\ref{eFFs} and~\ref{mFFs} we first show the electromagnetic nucleon form
factors as functions of the momentum transfer $Q^2$. The PFSM predictions are found
in surprisingly good agreement with the experimental data both for the proton and
the neutron~\cite{Wagenbrunn:2000es,Boffi:2001zb}.
This is essentially true for both the psGBE and OGE RCQMs. In addition,
the corresponding results compare well with the relativistic predictions by the
II RCQM from a Bethe-Salpeter approach. Only, the calculation with the wave function
from confinement only, which is spatially completely symmetric, fails drastically;
e.g., it yields an almost zero result for the neutron electric form factor (cf. the
right panel of fig.~\ref{eFFs}).

A nonrelativistic approach is by no means adequate for the electromagnetic form
factors~\cite{Wagenbrunn:2000es,Boffi:2001zb}. A comparison of the point-form with
the instant-form approach has revealed that in the spectator approximation for the
electromagnetic current, the instant form falls short too~\cite{Melde:2007zz}
(see figs. 5 and 6). In
particular, it was found that the instant-form spectator-model (IFSM) construction
is not frame-independent. This represents a considerable disadvantage as compared to
the PFSM, which strictly maintains its spectator-model character in all reference
frames and thus leads to manifestly covariant observables.
\begin{figure}[h]
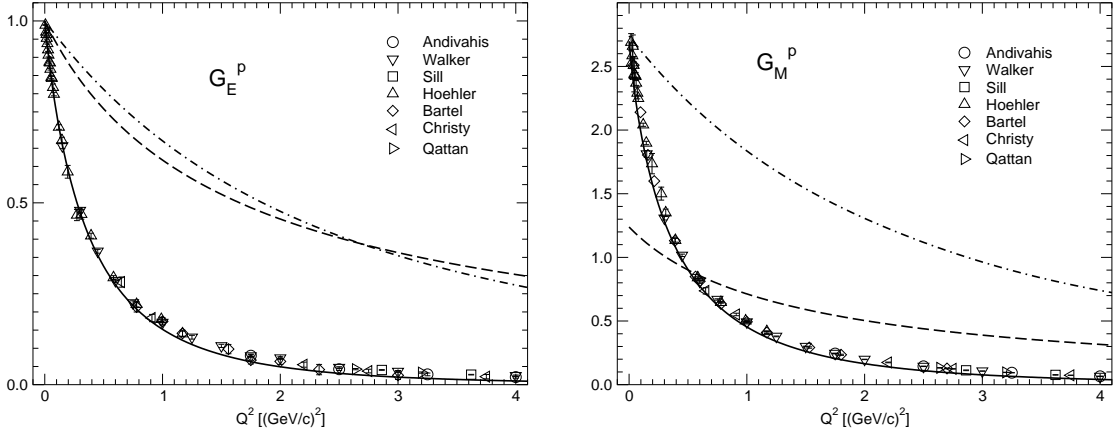

\label{pPFIF}
\begin{center}
\includegraphics[clip=,height=5.7cm]{gep_new.eps} \hspace{2mm}
\hspace{0.3cm}
\includegraphics[clip=,height=5.7cm]{gmp_new.eps}
\end{center}
\vspace{-3mm}
\caption{
Comparison of the electric and magnetic form factors of the proton as calculated with the
psGBE RCQM along the PFSM (full line) and the IFSM (dashed line). Also the nonrelativistic
impulse approximation is given (dash-dotted line). Experimental data as specified in the
inserts, see also ref.~\cite{Melde:2007zz}.}
\end{figure}
\begin{figure}[h]
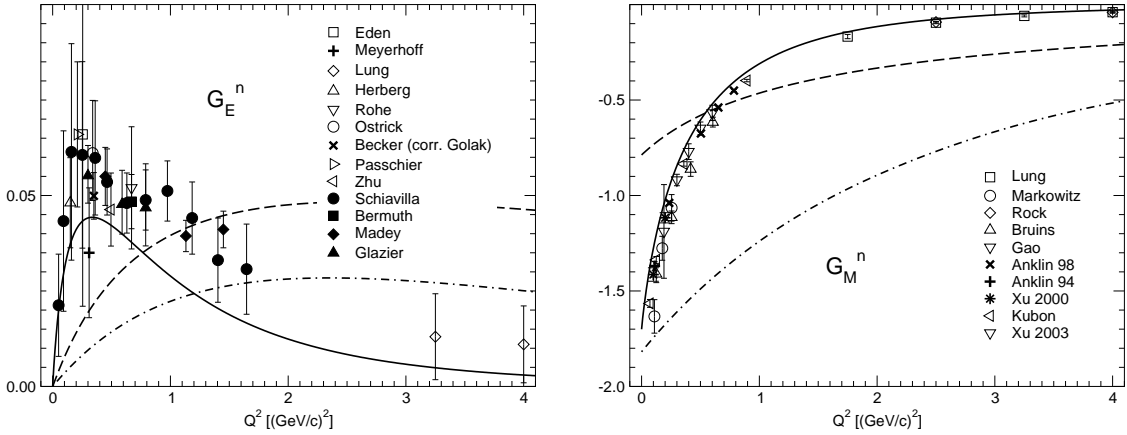

\label{nPFIF}
\begin{center}
\includegraphics[clip=,height=5.7cm]{gen_new.eps} \hspace{2mm}
\hspace{0.3cm}
\includegraphics[clip=,height=5.7cm]{gmn_new.eps}
\end{center}
\vspace{-3mm}
\caption{Same comparison as in fig. 5 but for the neutron.}
\end{figure}

In view of the relatively small extensions of the spatial probability distributions of
the nucleon wave functions in the rest frame, being centered around
$\sim$ 0.3 fm~\cite{Melde:2008yr},
one gets certainly curious, how the electric radius especially of the proton
(of $r_E\sim$ 0.877 fm) can be reproduced. Here the relativistic calculations along the
PFSM again yield very reasonable predictions~\cite{Berger:2004yi}
(see tab.~\ref{tab:chargeradii}). The same is true for the neutron and in case of
the $\Sigma^-$ (not quoted here), for the $r_E$ of which experimental data exist too.
Also the magnetic moments are at once described correctly, with only minor deviations
from phenomenology (see tab.~\ref{tab:magmom}). Again the IFSM results fail drastically
like the nonrelativistic impulse approximation (NRIA) does~\cite{Melde:2007zz}. 

\renewcommand{\arraystretch}{1.1}

\begin{table}[h]
\caption{\label{tab:chargeradii}Electric radii squared of the proton and neutron
(in fm$^2$) as predicted by the psGBE RCQM with the PFSM, IFSM, and the NRIA
current operators. Additional results (for all of the baryon ground states) can be found
in ref.~\cite{Berger:2004yi}. Experimental data after the PDG~\cite{Nakamura:2010zzi}.}
\begin{center}
\begin{tabular}{ccrrr}
Nucleon & Experiment & PFSM & IFSM & NRIA\\
\hline
 p & 0.769$\pm$0.012 & 0.824 & 0.156 & 0.102  \\
 n & -0.1161$\pm$0.0022 & -0.135 & -0.020 & -0.116 \\
\hline
\end{tabular}
\end{center}
\end{table}
\begin{table}[h]
\caption{\label{tab:magmom}Magnetic moments of the proton and neutron
(in n.m.) as predicted by the psGBE RCQM with the PFSM, IFSM, and the NRIA
current operators. Additional results (for all of the baryon ground states) can be found
in ref.~\cite{Berger:2004yi}. Experimental data after the PDG~\cite{Nakamura:2010zzi}.}
\begin{center}
\begin{tabular}{ccrrr}
Nucleon & Experiment & PFSM & IFSM & NRIA \\
\hline
 p & 2.792847356$\pm$0.000000023 & 2.70 & 1.24 & 2.74  \\
 n & -1.9130427$\pm$0.0000005 & -1.70 & -0.79 & -1.82 \\
\hline
\end{tabular}
\end{center}
\end{table}

\begin{table}[h]
\caption{Axial charges $g_A^B$ of octet and decuplet baryon ground states as predicted
by the EGBE RCQM~\cite{Glantschnig:2004mu} in comparison to
experiment~\cite{Nakamura:2010zzi} and lattice-QCD results from Lin and Orginos
(LO)~\cite{Lin:2007ap} and
Erkol, Oka, and Takahashi (EOT)~\cite{Erkol:2009ev} as well as results from
chiral perturbation theory by
Jiang and Tiburzi (JT)~\cite{Jiang:2008we,Jiang:2009sf}; also given is the
nonrelativistic limit (NR) from the EGBE RCQM.}
\begin{center}
\begin{tabular}{ccccccc}
Baryon & Experiment & EGBE  &LO&EOT&JT& NR\tabularnewline
\hline 
 {\it N} & 1.2694$\pm$0.0028 & 1.15 &  1.18$\pm$0.10&1.314$\pm$0.024&1.18& 1.65\tabularnewline
$\Sigma$ & $\cdots$ & 0.65   & 0.636$\pm$0.068$^\dagger$
&0.686$\pm$0.021$^\dagger$&0.73& 0.93\tabularnewline
$\Xi$ & $\cdots$  & -0.21 &  -0.277$\pm$0.034&-0.299$\pm$0.014$^\ddagger$&-0.23$^\ddagger$& -0.32\tabularnewline
\hline 
$\Delta$ & $\cdots$  & -4.48 & $\cdots$ & $\cdots$ &$\sim\,$-4.5& -6.00 \tabularnewline
$\Sigma^{*}$ & $\cdots$  & -1.06   & $\cdots$ & $\cdots$ & $\cdots$ & -1.41\tabularnewline
$\Xi^{*}$ & $\cdots$  & -0.75  & $\cdots$ & $\cdots$ & $\cdots$ & -1.00\tabularnewline
\hline
\end{tabular}
\end{center}
\begin{flushleft}
$^\dagger$ Because of another definition of $g_A^{\Sigma}$ this numerical value is
different by a $\sqrt{2}$ from the one quoted in the original paper.\\
$^\ddagger$ Because of another definition of $g_A^{\Xi}$ this value has a sign opposite
to the one in the original paper.
\end{flushleft}
\label{choi:t1}
\vspace{-6mm}
\end{table}

\begin{figure}[h]
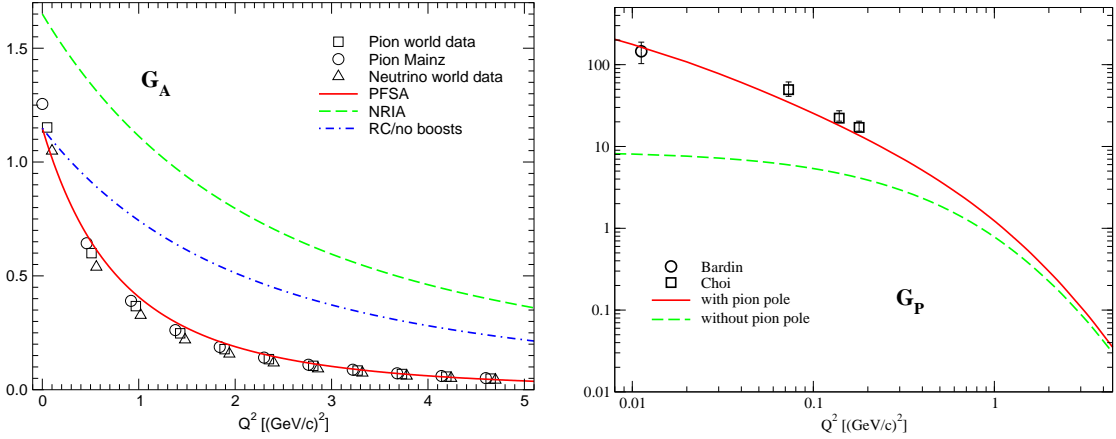

\begin{center}
\includegraphics[clip=,width=7.1cm]{ga_prd.eps} \hspace{4mm}
\includegraphics[clip=,width=7.1cm]{gp_prd.eps}
\end{center}
\vspace{-4mm}
\caption{Predictions of the psGBE RCQM for the axial (left) and induced pseudoscalar
form factors (right) of the nucleon as calculated along the PFSM (solid/red lines).
A comparison is given to the cases with a relativistic current but without boosts
(dash-dotted/blue line) and the NRIA (dashed/green lines). Experimental data as specified
in the inserts, see also refs.~\cite{Boffi:2001zb,Glozman:2001zc}.}
\label{aFFs}
\end{figure}

The axial nucleon form factors are also well described in the PFSM
approach~\cite{Glozman:2001zc,Boffi:2001zb} (see fig.~\ref{aFFs}). Similarly, the
axial charges of the nucleon as well as other baryon ground states turn out as
reasonable~\cite{Choi:2010ty} (see tab.~\ref{choi:t1}, where the predictions of
the EGBE RCQM are quoted).
A study of the axial charges of $N^*$ and other baryon
resonances has in addition revealed that the predictions of the GBE RCQM agree well 
with data so far known from lattice QCD~\cite{Choi:2010ty,Choi:2009pw}. 

\section{Structure of meson-baryon interaction vertices}

\begin{figure}[h]
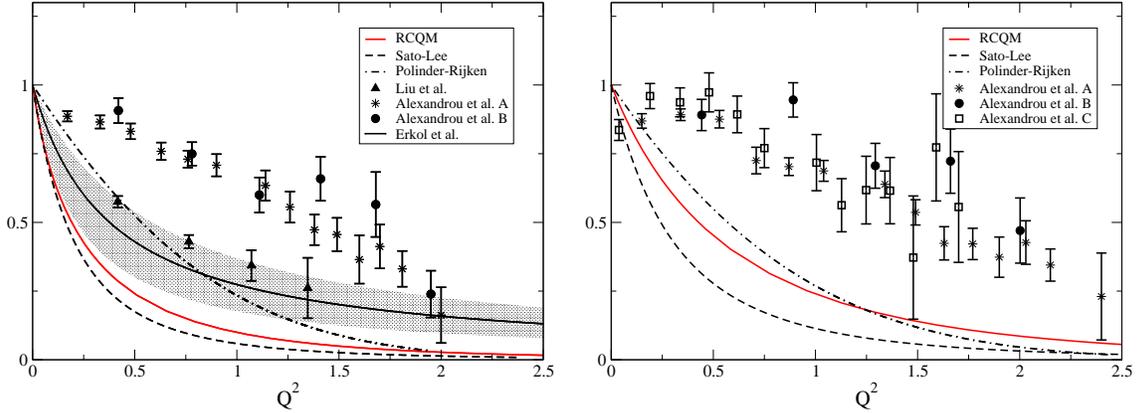

\begin{center}
\includegraphics[clip=,width=7.3cm]{nucleon.eps} \hspace{2mm}
\includegraphics[clip=,width=7.3cm]{delta.eps}
\end{center}
\vspace{-4mm}
\caption{Predictions of the strong vertex form factors $G_{\pi NN}$ (left) and
$G_{\pi N\Delta}$ (right) by the psGBE RCQM (solid/red line) in comparison to
parametrizations from the dynamical meson-baryon
models of Sato-Lee~\cite{Sato:1996gk} and Polinder-Rijken~\cite{Polinder:2005sm,
Polinder:2005sn} as well as results from lattice QCD
calculations~\cite{Alexandrou:2007zz,Liu:1994dr,Liu:1998um,Erkol:2008yj} (cf. the
legend); in the left panel the shaded area around the result by Erkol {\it et al.}
(thick solid line) gives their theoretical error band. For more details see
ref.~\cite{Melde:2008dg}.}
\label{vFFs}
\end{figure}

The baryon axial properties are closely related to the meson-baryon couplings through
the Goldberger-Treiman relation. Therefore, on has recently also studied the predictions
of the psGBE RCQM for the structures of the $\pi NN$ as well as $\pi N\Delta$
interaction vertices~\cite{Melde:2008dg}. It has turned out that the relativistic
microscopic theory relying on the meson-quark coupling provided by the GBE dynamics
yields quite reasonable descriptions of the momentum dependences of these strong form
factors, in the sense that previous parameterizations within phenomenological 
meson-nucleon or meson-$\Delta$
models~\cite{Sato:1996gk,Polinder:2005sm,Polinder:2005sn} have arrived at rather similar
results (see fig.~\ref{vFFs}). Of course, the strong vertex form factors $G_{\pi NN}$
and $G_{\pi N\Delta}$ are not experimentally observable, and a comparison is possible only
to other theoretical approaches. Lattice-QCD results exist from various groups but,
unfortunately, they do not yield a unique picture. In view of the considerable differences
of the RCQM predictions with the lattice-QCD results and of the lattice-QCD results among
themselves, it would certainly be desirable to have a more reliable data base for the
comparison. In any case, it should be noted that the magnitude of the $\pi NN$ coupling
constant $f_{\pi NN}$ extracted from the vertex form factors of the GBE RCQM is not only
in good agreement with the phenomenological value of 0.075 but also fits to the nucleon
axial charge through the Goldberger-Treiman relation~\cite{Melde:2008dg}.

\section{Concluding remarks}

The point form has not been much used until about a decade ago. Once first results on
elastic $N$ form factors have become available around the year 2000,
they appeared as rather surprising and could be seen as
incidentally in agreement with phenomenology. However, with the advent of ever more 
predictions derived along the point form, this approach has become truly remarkable,
especially since the very reasons for the good performance are not yet fully understood.
Further studies are thus strongly advised, above all consistent
comparisons with the front and instant forms.

\noindent
{\bf Acknowledgement} \\[2mm]
I am grateful to the organizers of the LC2010 conference for having given me the opportunity
to present this review of recent achievements within point-form relativistic quantum
mechanics at a meeting that primarily focussed on light-front quantum mechanics and
light-front field theory.\\
The results discussed here have arisen over several years with essential contributions
by K.~Berger, S.~Boffi, L.~Canton, K.-S.~Choi, B.~Desplanques, K.~Glantschnig, L.~Y.~Glozman, R.~Kainhofer, W.~Klink, T.~Melde, Z.~Papp, M.~Radici, B.~Sengl, L.~Theussl, 
K.~Varga, and R.~F.~Wagenbrunn.

\end{document}